\documentclass[letterpaper]{jpconf}
\usepackage{graphicx}
\usepackage{amsmath}
\usepackage{xcolor}
\begin{document}
\title{Topological spin-torque in non-collinear anti-ferromagnetic 3Q state}
\author{Adel Abbout}

\address{King Fahd University of Petroleum and Minerals, Dhahran, Saudi Arabia}

\ead{adel.abbout@kfupm.edu.sa}

\begin{abstract}
We investigate the spin torque in the topological phase of the 3Q antiferromagnetic (AFM) configuration. We first obtain the band structure to identify the topology and nature of the different gaps of the system and then calculate the spin density in the whole system. We demonstrate the presence of a non-vanishing spin-torque component at the edge of the system and analyze its effect on the texture. Moreover, we show that the direction of the torque depends hugely on the way we cut the system to define the edge.

\end{abstract}
\section{Introduction}

The race behind low-energy consumption devices led to the investigation of a multitude of systems exhibiting topological behaviour \cite{Bonbien_2022,MacDonald}. Indeed, the presence of states robust against disorder and deformations with quantized topological properties opens the door to outstanding features \cite{MacDonald}. One of the key elements in topological systems is spin-orbit coupling and symmetry breaking. This interaction might lead in some systems to the emergence of a non-vanishing Berry curvature and the presence of an anomalous Hall  effect \cite{Bonbien_2022,Papa_2019}. Recently, the investigation of non-collinear \cite{Nakatsuji2015,Nayak2016} and non-coplanar anti-ferromagnets \cite{Martin2008} showed that a topological state can be obtained without the presence of spin-orbit coupling \cite{Chen2014,CKubler2014}. As a matter of fact, the 3Q state that appears in Mn/Cu(111) interface \cite{Kurz}, presents an interesting non-collinear and non-coplanar texture with 4-spin unit cell in a triangular lattice with the spins having the same angle between them. \textcolor{black}{This non-conventional texture is the reslut of a phase transition from a coplanar row-wise AFM configuration \cite{Kurz,Jolicoeur_1990} to a non-coplanar one due to the presence of 4-spin exchange interaction \cite{Kurz,Blgel2005,Blugel}. Indeed, \textit{ab initio} calculation with a Heisenberg model show that this interaction induces a ground state where the 4-spins of the unit cell are living in 3D while keeping their sum to be zero. }
In a finite-size sample, the spins at the edge might have a different orientation than those of the bulk because of the absence of the effective field coming from the missing neighbours. Hence, these local spins are expected to  feel a  larger spin torque from the electronic spin density. This effect is more likely to happen in the case of a state in the topological phase since all the 
\begin{figure}[h]
\begin{center}
\includegraphics{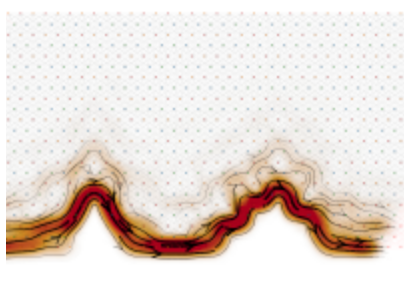}
\end{center}
\caption{\label{EdgeCurrent}Sketch of the  edge charge current in the topological phase of a magnetic system. The current is following the deformation, and propagates along the edge. }
\end{figure}
charge carriers propagate at the edge of the sample as shown in Figure \ref{EdgeCurrent}. 

Investigating the effect of edge spin density and its corresponding spin torque on the magnetic texture is very important to specify whether the texture can remain stable or will collapse to another quasi-ground state. An interesting approach to a more realistic investigation would be to run an atomistic simulation and generate the exact texture of the ground state \cite{Evans_2014}. In parallel, a tight-binding simulation will provide the electronic spin density that can be incorporated in the effective field of the atomistic simulation.
In this study, we will focus on the tight-binding simulation to express the electronic spin densities and their spin torque on the texture in the 3Q state configuration.

\section{Model}
\begin{figure}[h]
\begin{minipage}{14pc}
\includegraphics[width=15pc]{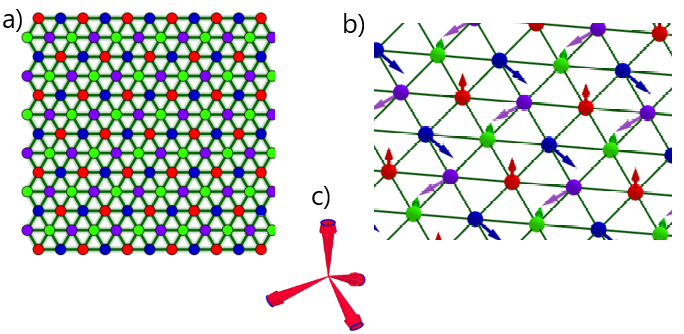}
\caption{\label{system}Lattice model for the 3Q state. a) The triangular lattice with 4 sites unit cell (the next nearest neighbours hopping are not depicted for clarity of the figure). b) The local spins configuration. All the spins are non-collinear and do not live on the same plane. c) The four different local spins of a unit cell plotted from the same origin. The angles between each two local spins are the same. For the transport properties, the system will be considered infinite in the x-direction while the width is kept fixed. }
\end{minipage}\hspace{6pc}%
\begin{minipage}{14pc}
\begin{center}
\includegraphics[width=9pc]{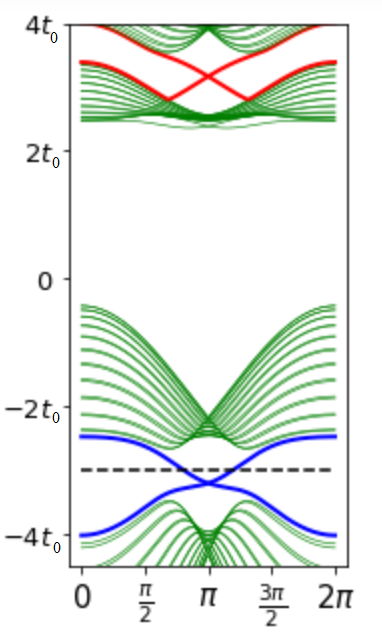}
\caption{\label{Band}The band strucure for non-collinear antiferromanetic 3Q texture. The next nearest hopping is $t^\prime=0.3t_0$. The blue and red bands represent the two edge states corresponding to the two topological gaps. The dashed line shows the Fermi level $E_F=-3t_0 $. The exchange parameter is $\Delta=3t_0$}
\end{center}
\end{minipage} 
\end{figure}

The antiferromagnetic 3Q texture consists of localized spins on a triangular lattice with a spin Hamiltonian $\mathcal{H}=-J_1\sum_{\langle i,j\rangle} \mathbf{S}_i \mathbf{S}_j-J_2 \sum_{\langle\langle i,j\rangle\rangle} \mathbf{S}_i \mathbf{S}_j$. $J_1$ and $J_2$ are antiferromagnetic exchange coefficients with nearest and next-nearest neighbours interactions respectively \cite{Jolicoeur_1990}. This Hamiltonian is rich in its phase space and one needs to add the 4-spin interaction to stabilize the 3Q state with its 4-spins having the same angle between them ($\theta_0\approx 109.47^\circ$) as shown in Fig. \ref{system} \cite{Kurz,Blugel}. The 4 spins making the unit cell are non-collinear and non-coplanar, with a sum $\sum_{i\in \textrm{cell}}\mathbf{S}_i=0$ \textcolor{black}{\cite{Papa_2019,Martin2008,Blugel}}. Unlike a conventional planar antiferromagnetic texture, the chirality $\kappa=\mathbf{S}_k .(\mathbf{S}_i\times \mathbf{S}_j )\ne 0$ which provides an effective spin-orbit coupling for the electrons hopping in the lattice \textcolor{black}{\cite{Papa_2019,Chen2014}}. Consequently, a non-zero-Berry curvature is found and the signature of quantum-anomalous Hall  effect is seen in the $\sigma_{xy}$ conductivity of electrons. 
The Hamiltonian describing electrons in the system tightly coupled to the local spins with sd interaction can be described in the tight-binding approximation as \cite{Martin2008,Adel}:
\begin{equation}
    H=-\sum_{i,j} t_{ij} \hat{c}_i^\dagger \hat{c}_j+\Delta \sum_i \hat{c}_i^\dagger (\mathbf{S}_i .\hat{\boldsymbol{\sigma}})\hat{c}_i
\end{equation}
with the first sum is done over the first ($t_{ij}=t_0$) and second ($t_{ij}=t'$) nearest neighbours only. $\Delta$ is the s-d interaction exchange and  $\hat{\boldsymbol{\sigma}}$ is the Pauli matrices vector. $\hat{c_i}^\dagger$ ($\hat{c}_i$) is the creation (annihilation) operator for electrons on site $i$. (The electron spin index is hidden in the operator ${\hat{c}_i\equiv (\hat{c}_{i\uparrow},\hat{c}_{i\downarrow})}$). The lattice is shown in Figure \ref{system} as well as the non-coplanar magnetic structure with an antiferromagnetic unit cell.  The band structure for relatively large $\Delta $ (compared to the hopping parameter $t$) exhibits two topological gaps and a trivial one, with bands being two-fold degenerate \cite{Papa_2019}. As shown in Figure \ref{Band}, the band for a semi-infinite quasi-1D waveguide shows two edge states  at the top and bottom gaps, revealing a signature of a topological behaviour of the system for the corresponding set of energies. These edge states are non-degenerate and provide a quantum of conductance $\frac{e^2}{h}$ in the  conductance profile. In the case of an infinite system, the topology appears in the presence of a quantum anomalous Hall  effect, manifesting as a quantum conductance in the transverse conductance measurement $\sigma_{xy}$.
From the atomistic Landau-Lifshitz-Gilbert simulation point of view, one can construct the structure using the magnetic Hamiltonian described previously and calculate the effective fields to stabilize iteratively the structure towards the ground state \cite{Evans_2014}. In this process, the use of periodic boundary condition may speed up the stabilization process. In a real situation, the conditions at the edge are open, and the local spins are missing half of the contribution of the effective field coming from the missing neighbours at the edge. For this reason, one expects the local spins at the edge to deviate from the equilibrium direction of the spins of the bulk. In the presence of quantum transport, for energies in the toplogical gap, the edge states build up an electronic spin density at the edge that couples to the local atomic spins and exert a spin torque on them. The exerted spin torque will challenge the local equilibrium of the spins and can even destroy the non-coplanar phase and thus the topological (electronic and magnetic) properties of the system. For this reason, it is important to study the spin density and the spin torque in such system.

To investigate the spin torque, we consider a system with next nearest neighbours $t^\prime=0.3t_0$ and an exchange strength $\Delta=3 t_0$ . The Fermi energy is set to $E_F=-3 t_0$ to allow one single topological mode at the bottom gap. The spherical angles $(\theta,\phi)$ defining the local spins in the unit cell are $(\theta_0,-\frac{\pi}{3})$, $(\theta_0,\frac{\pi}{3})$,$(0,0)$,$(\theta_0,\pi)$, where $\theta_0=\arccos(-\frac{1}{3})$. The system is infinite in the x-direction and has a width $W=20a$, $a$ being the lattice constant. We use the Kwant package to calculate the wave function and the spin densities on each site of the system \cite{KWANT}.
\section{Results}
Figure \ref{system} shows the studied lattice system with the position of the local moments on the sites and the angles between the spins. The band structure for the quasi 1D system is depicted in figure \ref{Band}. The middle gap is trivial whereas the top and bottom ones are topological, recognized by the presence of two edge states (shown in red and blue) \cite{Papa_2019}.

\begin{figure}
\begin{center}
\includegraphics[width=35pc]{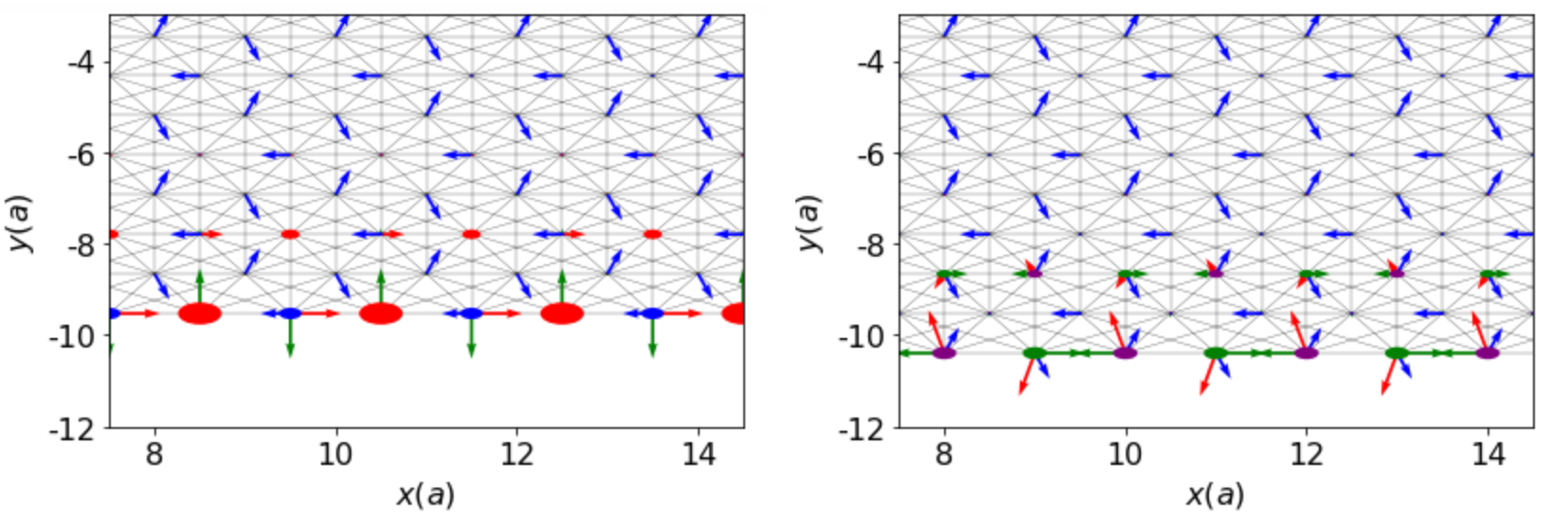}
\end{center}
\caption{\label{torque} The lattice model of the 3Q antiferromagnetic state. The figures show the lower edge of a uniform waveguide infinite in the x-direction.  The blue arrows represent the in-plane local moments (spins).  The colours for the unit cell sites are green, purple, blue and red for the corresponding spherical angles $(\theta_0,-\frac{\pi}{3})$, $(\theta_0,\frac{\pi}{3})$, $(0,0)$, $(\theta_0,\pi)$ respectively. The difference between the left and right figure is in the sites lying at  the lower edge of the sample. The size of the sites is proportional to the $ \langle s_z\rangle$  component of the electron spin density. The in-plane electron spin density $ \langle \boldsymbol{s}_\|\rangle$ is depicted with a red arrow. The in-plane spin torque $\boldsymbol{\tau}_\|$ is depicted  with a green vector.
(In these figures, "a" is the lattice parameter)}
\end{figure}

By choosing the Fermi energy in the middle of the lower gap ($E_F=-3t_0$), we select the edge state that propagates at the lower edge of the system. We can see in Figure \ref{torque} that the state lives on the edge. Indeed, the size of each site is put to be proportional to the out-of-plane electron spin density and it clearly shows that the state barely penetrates two unit cells and decays exponentially from the lower edge of the system. The local moments at the edge can be different from one sample to another as it can be seen in Figures \ref{torque} left and right. In fact, depending on the direction of the cut of the system, we can have any couple of the spherical angles defined previously. The direction of the spin torque depends hugely on the sites lying on the edge since the built-up electronic spin density depends on the magnetization at that site. The in-plane spin density in Figure \ref{torque} left is opposite to the local magnetization due to the positive sign of the exchange parameter $\Delta$. This direction of the spin-density is not due solely to the sign of the exchange parameter as it can be deduced from Figure \ref{torque} (right). Indeed, a tilt is visible between the spin density (red arrows) and the local moments (blue arrows). Since the electron is hopping between sites of different local moments, one can not predict the direction of the spin-density by hands, without proper calculation, despite that the sd exchange interaction is antiferromagnetic in our model. 
The spin torque reflects the effect of the spin density on the magnetic structure and how it tends to modify it. The most straightforward and natural way to define it is \cite{Akosa_2017,Papa_2017}:

\begin{equation}
\label{torqueeq}
\boldsymbol{\tau_i}=\frac{2\Delta}{\hbar}\langle \boldsymbol{s_i}\rangle\times \boldsymbol{S_i}    
\end{equation}
where $i$ refers to the site $i$. The spin density at the Fermi level  is evaluated using the edge state following the formula:
\begin{equation}
    \langle\boldsymbol{s_i}\rangle=\boldsymbol{\psi}_i^{\dagger} \hat{\boldsymbol{\sigma}} \boldsymbol{\psi}_i
\end{equation}
From equation \ref{torqueeq}, we deduce that the spin torque in the bulk of the system is exponentially vanishing and only survives near the edge of the system \textcolor{black}{(because so is the wave function)}. This is confirmed by Figures \ref{torque} where the green arrows showing the  spin torque are mainly on one or two unit cells from the edge. A non-vanishing torque at the edge means that the local moments at the edge will precess to a new equilibrium state in the presence of damping \cite{Evans_2014}. This new configuration at the edge will alter the equilibrium of spins in the next layer from the edge and we can continue with the same reasoning for the inner layers until we get deep into the bulk. This change in the magnetic structure can be negligeable  and manifests as a small tilt in the local moments without destroying the main transport features of the system. On the other hand, when the magnitude of the spin torque is large enough, the whole magnetic structure starts collapsing and a change of phase will occur with completely different magnetic and transport properties \cite{Durga_2021}. 
In order to have a clear picture of the expected magnetic structure, one needs to perform atomistic simulation using Landau-Lifshitz-Gilbert equation  to predict the stable ground state for the given density of states \cite{Evans_2014}. Moreover, we need to make sure to keep open boundary conditions since the main effect we are investigating comes from the presence of the edge.
\section{Conclusion}
The spin density that builds up at the edge of the topological phase of the 3Q antiferomagnetic state does not follow the direction of the local moments and thus exerts a torque on the magnetic structure. The direction of the torque depends on the local moments present at the edge and might vary from one sample to another. The fact that in the topological phase, the edge state is mainly concentrated on a tiny width from the edge, the involved spin densities are expected to be large enough to be accounted in the effective field inducing  the equilibrium of the magnetic ground state. \textcolor{black}{ Consequently, the stability of the 3Q state depends on the the polarization of the current  injected from the leads which can sustain a important spin density on the edge. The non-vanishing exerted torque will most likely modify the anti-ferromagnetic order at the edge and change the local transport properties of the system.}
\section*{Aknowledgement}
The author gratefully
acknowledges the support provided by the Deanship of
Research Oversight and Coordination (DROC) at King
Fahd University of Petroleum and Minerals (KFUPM)
for funding this work through exploratory research grant
No. ER221002.

\section*{References}

\bibliographystyle{iopart-num}

\bibliography{reference}
\end{document}